\documentclass[10pt]{aa}
\usepackage{graphicx}
\usepackage{txfonts}
\usepackage{natbib}
\def\teff{$\mathrm T\rm_{eff }$}
\def\kms{$\mathrm {km s}^{-1}$}
\begin{document}
   \title{UVES observations of the Canis Major overdensity
   \thanks{Based on observations obtained in the ESO Director's Discretionary Time program 272-B.5017}}


   \author{L. Sbordone
          \inst{1,3}
          \and
          P. Bonifacio\inst{2}
          \and
	  G. Marconi\inst{1}
	  \and
	  S. Zaggia\inst{2}
	  \and
	  R. Buonanno\inst{3}
          }

   \offprints{L. Sbordone}

   \institute{ESO - European Southern Observatory, Alonso de Cordova 3107, Vitacura, Santiago de Chile
         \email{lsbordon@eso.org}
         \and
             INAF - Osservatorio Astronomico di Trieste, via G. B. Tiepolo, Trieste ITALY
	 \and
             Universit\'a di Roma ``Tor Vergata'', via Della Ricerca Scientifica 2, Roma ITALY
             }
\authorrunning{Sbordone et al.}
\titlerunning{Canis Major}

   \date{Received ; accepted }

   \abstract{

We  present the first   detailed  chemical abundances for three  giant
stars which are candidate members of  the Canis Major overdensity, obtained by
using FLAMES-UVES at VLT. The  stars, in the background
of the open cluster \object{NGC 2477}, have radial velocities compatible
with a membership to this structure. However,  due to  Galactic disc contamination, radial  velocity by  itself  is  unable to firmly
establish membership.  The metallicities span  the
range $\rm -0.5\la [Fe/H] \la +0.1$. Assuming that at least one of
the  three stars is  indeed  a member of CMa implies that this
structure   has undergone  a   high   level of  chemical   processing,
comparable to that of the Galactic  disc.  The most metal-rich
star of the sample, \object{EIS 6631},  displays several abundance ratios which
are  remarkably     different  from    those  of   Galactic     stars:
[$\alpha$/Fe]$\sim   -0.2$, [Cu/Fe]$\sim  +0.25$, [La/Fe]$\sim  +0.6$,
[Ce/Fe]$\sim +0.8$ and [Nd/Fe]$\sim +0.6$. These ratios make it likely
that this star was formed in an external galaxy.

   \keywords{ Stars: abundances   -- Stars: atmospheres --   Galaxies:
   abundances -- Galaxies: evolution -- Galaxies: dwarf } }

   \maketitle
%

\section{Introduction}

In the framework  of the hierarchical  merging scenario for the galaxy
formation, dwarf galaxies play the role of ``building blocks'' of
the larger  structures like the   Milky        Way
(MW). Nevertheless, the \emph{present day} dwarf galaxies in the Local
Group (LG)  appear to be somewhat undesirable  building blocks:
their  chemistry is significantly different from the one found both in the MW Disc and Halo systems
\citep{venn04}. This is not surprising, since a long evolution took place,
after the main  merging phase, in the
``survived''     dwarf   galaxies  \citep{lanfranchi03}. Nevertheless,
merging  events  are  still taking place   in the  MW,   as
testified by  the discovery in the Halo  of the  stream related to the
Sagittarius    dwarf    Spheroidal      galaxy  \citep[Sgr dSph, see][]{ibata01,majewski03}.
The  Sgr   dSph  itself  displays   a  peculiar
chemical composition \citep{bonifacio00,bonifacio04},
leading to think that chemically peculiar subpopulations, traces of
past or ongoing merging events, should be identifiable in the Galactic
Disc or Halo.

Recently \citet{martin03} claimed the discovery of the  core of  a  tidally  disrupted dwarf  galaxy,   still
recognizable as an overdensity in the external  Galactic disc in Canis
Major (Canis   Major Overdensity or CMa from now on).
In a  subsequent paper \citep{bellazzini04} they recognized
the  same  population also   in the  background  of the  Galactic open
cluster \object{NGC 2477}  at $\simeq13^\circ$ from  the CMa centre.   The authors  situate the structure at  about 7 Kpc  from the
Sun and about 16 Kpc from the Galactic  Centre, and estimate a mass of
about  $10^{7}~M_{\sun}$,  which   would make it   the  nearest known
external galaxy.   They also associate it to  the
ring-like structure known as Monoceros Ring \citep{newberg},
Ring \citep{ibata} or GASS (Galactic Anticentre Stellar Structure, see
\citealt{crane03} and \citealt{frinchaboy04}).
\citet{bellazzini04} also inferred a possible connection with some Galactic globular
clusters, among others \object{NGC 2808}.  Shortly afterwards, \citet{momany04}
questioned the  effective existence  and size  of CMa,   claiming that 
the  anomaly could be
explained, to  a  large extent,      
by properly taking into  account the
Galactic disc warp, which is maximum in the   
CMa direction.  \citet{bellazzini04} examined and rejected
this hypothesis, and so 
did \citet{martin04}, deriving for the centre of the 
structure a
radial velocity of 109 \kms, with a low
velocity dispersion of 13 \kms, both difficult to
reconcile with the dynamics of the local disc.


\begin{table*}[t]
{
\begin{tabular}{cc}
\begin{minipage}{4cm}
\vspace*{-1.2cm}
\caption{Photometry and physical parameters for the three stars.}
\end{minipage}
\hspace*{0.1cm}
\begin{tabular}{lccrrrrrrr}
\hline
{\bf Star} & \multicolumn{2}{c}{$\alpha$\,\, (J2000.0)\,\, $\delta$} &
{\bf V} & {\bf (V-I)}{\boldmath $_0$} &
{\bf T}{\boldmath $\rm_{eff}$} & {\boldmath $\log{g}$} & [Fe/H] & {\boldmath $\xi$} &{\boldmath $\rm V \rm_{rad}$} \\
 & hms &$^\circ,',''$ & mag  & mag & K  & cgs & dex  & \kms & \kms \\
\hline
\object{EIS 6631}  & $07\, 51\, 36.2$ &$-38\, 31\, 10$ & 16.35 & 0.75 & 5367 & 3.5 & +0.15 & 1.80 & 135.4\\
\object{EIS 7873}  & $07\, 52\, 37.1$ &$-38\, 28\, 01$ & 16.26 & 0.89 & 4990 & 2.3 & --0.42 & 1.80 & 111.1\\
\object{EIS 30077} & $07\, 51\, 41.0$ &$-38\, 38\, 39$ & 16.51 & 0.89 & 4994 & 2.8 & --0.04 & 1.45 & 97.0 \\
\hline
\end{tabular}
      \end{tabular}
}
\label{tabuno}
\end{table*}


\section{Data reduction and analysis}

Shortly after the announcement of the discovery of CMa, we obtained
Director's Discretionary  Time (DDT) at   VLT-FLAMES with  the aim  of
probing the dynamics and  chemical composition of the newly  discovered
structure.
\citet{bellazzini04} detected the CMa population in the background of
\object{NGC 2477} using the EIS pre-FLAMES photometry and astrometry of
\citet{momany01} which is publicly available.
We therefore observed this  field selecting in
the EIS photometry Red Giant/Clump stars
possible CMa  members.
Observations were performed between January and   March  2004
and consisted  of
$4\times3045$  seconds exposures, using the   HR09
setting for GIRAFFE
fibers, and the UVES setting centred at  580 nm.

\begin{figure}
   \centering
   \includegraphics[width=6.5 cm]{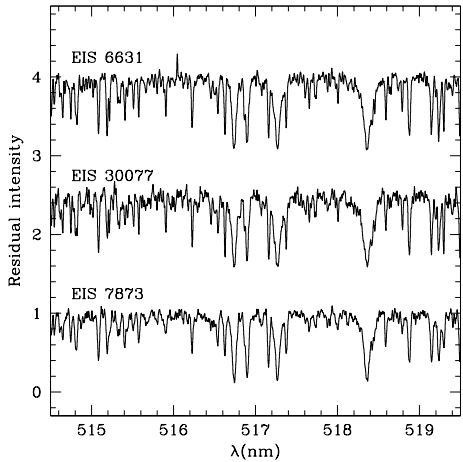}
      \caption{Spectra  of the three most probable CMa stars,
               in the region of the Mg b triplet.
               [Fe/H], $\log(g)$ and \teff\ all increase from
               bottom to top. The spectra are normalized to one, stars
	       30077 and 6631 are shifted vertically for display purposes (continuum is at 2.5 for 30077 and at 4 for 6631). }
         \label{figspe}
   \end{figure}

\begin{table}[t]
\begin{center}
\caption{Abundance ratios for the three stars. [X/\ion{Fe}{i}] is used for neutral elements, [X/\ion{Fe}{ii}] for ionized
species, and [Fe/H] for \ion{Fe}{i} and \ion{Fe}{ii}. Errors are $1~\sigma$ intervals, ``n'' is the number of lines used}
{\scriptsize
\begin{tabular}{rrrrrrr}
\hline
         & {\bf \object{EIS 6631}}&          & {\bf \object{EIS 7873}} &         & {\bf \object{EIS 30077}} &         \\
         & [X/Fe]        &  n        & [X/Fe]         & n        & [X/Fe]          & n        \\
\hline
\ion{Na}{i}     &$0.18 $$\pm 0.18$ &4          &$  0.18$$ \pm 0.10 $  &3        &$ -0.10$$ \pm 0.15 $ &4         \\
\ion{Mg}{i}     &$-0.49$$\pm 0.11$ &3          &$ -0.18$$ \pm 0.23 $  &3        &$ -0.01$$ \pm 0.15 $ &4         \\
\ion{Al}{i}     &$-0.21$$\pm 0.14$ &2          &$ -0.14$$ \pm 0.13 $  &2        &$ -0.24$$ \pm 0.14 $ &2         \\
\ion{Si}{i}     &$-0.21$$\pm 0.14$ &4          &$  0.10$$ \pm 0.11 $  &4        &$ -0.27$$ \pm 0.16 $ &4         \\
\ion{Ca}{i}     &$-0.22$$\pm 0.16$ &11         &$  0.02$$ \pm 0.16 $  &9        &$ -0.08$$ \pm 0.18 $ &7         \\
\ion{Sc}{ii}    &$-0.03$$\phantom{\pm 0.00}$&1 &$ -0.28$$ \pm 0.14 $  &2        &$ -0.20$$ \pm 0.20 $ &2         \\
\ion{Ti}{i}     &$0.18 $$\pm 0.16$ &7          &$ -0.02$$ \pm 0.14 $  &5        &$  0.13$$ \pm 0.16 $ &7         \\
\ion{V}{i}      &$0.33 $$\pm 0.19$ &3          &$  0.31$$ \pm 0.12 $  &2        &$  0.49$$ \pm 0.14 $ &2         \\
\ion{Mn}{i}     &$-0.02$$\phantom{\pm 0.00}$&1 &$ -0.12$$ \phantom{\pm 0.00}$ &1&$  0.12$$ \phantom{\pm 0.00}$ &1\\
\ion{Fe}{i}     &$0.15 $$\pm 0.11$ &20         &$ -0.42$$ \pm 0.10 $  &15       &$ -0.04$$ \pm 0.13 $ &21        \\
\ion{Fe}{ii}    &$0.12 $$\pm 0.17$ &13         &$ -0.37$$ \pm 0.13 $  &8        &$ -0.07$$ \pm 0.08 $ &8         \\
\ion{Co}{i}     &$0.25 $$\pm 0.21$ &2          &$ -0.17$$ \phantom{\pm 0.00}$ &2&$  0.06$$ \pm 0.13 $ &2         \\
\ion{Ni}{i}     &$0.02 $$\pm 0.18$ &15         &$ -0.09$$ \pm 0.19 $  &13       &$ -0.21$$ \pm 0.15 $ &11        \\
\ion{Cu}{i}     &$0.25 $$\phantom{\pm 0.00}$ &1&        $          $  &         &$  0.23$$ \phantom{\pm 0.00}$ &1\\
\ion{Y}{ii}     &$0.19 $$\pm 0.21$ &3          &$ -0.15$$ \pm 0.13 $  &2        &$ -0.61$$ \pm 0.13 $ &3         \\
\ion{Ba}{ii}    &$0.21 $$\phantom{\pm 0.00}$ &1&$  0.37$$ \phantom{\pm 0.00}$ &1&$  0.07$$ \phantom{\pm 0.00}$ &1\\
\ion{La}{ii}    &$0.61 $$\pm 0.24$ &2          &$  0.26$$ \pm 0.14 $  &2        &$  0.67$$ \pm 0.10 $ &2         \\
\ion{Ce}{ii}    &$0.78 $$\pm 0.25$ &3          &$  0.16$$ \phantom{\pm 0.00}$ &1&$  0.21$$ \pm 0.09 $ &2         \\
\ion{Nd}{ii}    &$0.59 $$\pm 0.20$ &2          &$ -0.01$$ \pm 0.18 $  &3        &$  0.36$$ \pm 0.13 $ &4         \\
\ion{Eu}{ii}    &$0.21 $$\phantom{\pm 0.00}$ &1&$  0.20$$ \phantom{\pm 0.00}$ &1&$  0.10$$ \phantom{\pm 0.00}$ &1\\
\hline      
\end{tabular}
}
\label{tabdue}
\end{center}
\end{table}

In this letter we describe the detailed chemical analysis  of 3 of the
7 UVES stars obtained with FLAMES; the analysis  of the stars observed
with GIRAFFE has been  described briefly in \citet{zaggia04}  and will
be the object of a separate paper.  The  four  spectra of each star
have been corrected to heliocentric  radial velocity and then coadded.
Due to the very low S/N ratio, they have  been convolved with a 5 \kms\
FWHM gaussian, degrading the resolution to about 33000, reaching a S/N
of about  40 per  pixel at 580  nm.
By  combining  our UVES and
GIRAFFE radial velocity measurements with those of \citet{martin04} in
the direction of CMa centre, those  of \citet{yanny} for the Monoceros
Ring  and  data for    open  and  globular clusters,  we    show  in
\citet{zaggia04} that,  assuming all these  objects belong to the same
structure, its motion is consistent with an  object in circular motion
at  the distance  of 15 kpc  from the   Galactic Centre,  and circular
velocity of 220 \kms.   This result  is not  very different  from that
derived by \citet{frinchaboy04} who prefer a distance of 18 kpc.  In the observed field CMa  member  stars are thus expected to
have   radial velocities $>100$ \kms.   This  led us to
exclude from our sample two stars (EIS  4383 and EIS 31581) which
showed  near-to-zero radial velocities.   E(V-I)  color excesses  were
derived from \citet{schlegel98} maps, as corrected by
\citet{bonifacio00b}.
Effective  temperatures  were derived  from the
\citet{alonso99} calibration for giant stars.
The abundance analysis was performed in a  traditional manner by using
our  Linux porting   of  the ATLAS,   WIDTH   and  SYNTHE codes  (see
\citealt{kurucz93} and  \citealt{sbordone04}).   We noticed  that,  at
variance with the other two stars, the lines  of \object{EIS 7873} appear to be
somewhat broader than the instrumental resolution.
We   derived the  final
gravity  by forcing \ion{Fe}{i}  --  \ion{Fe}{ii}   ionization equilibrium. In  this
phase, two more stars (EIS 2812 and EIS 5429) proved  to be
dwarfs ($\log g>4.0$),and thus
incompatible with a heliocentric
distance of the order of 7 kpc.  The coordinates,
photometry and atmospheric  parameters  for the three  remaining stars
are detailed in Table \ref{tabuno}, which provides also
heliocentric  radial velocities.

Synthetic  spectra
were computed with  SYNTHE  to derive the elemental  abundance, taking
into account HFS splitting for \ion{Co}{i}, \ion{Cu}{i}, \ion{Eu}{ii}.

   \begin{figure}
   \centering
   \includegraphics[width=6.5cm]{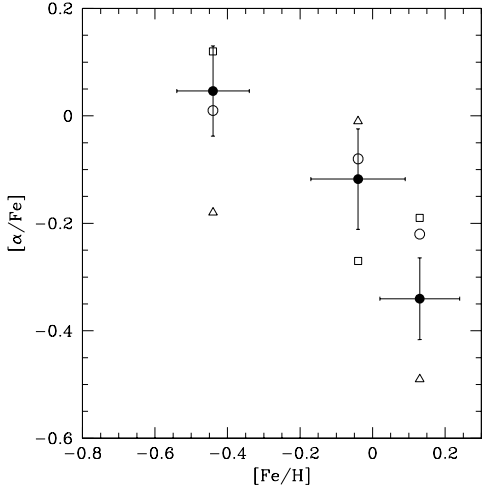}
      \caption{ Alpha elements abundances for the three sample stars,
      plotted against [Fe/H]. Triangles are \ion{Mg}{i}, squares \ion{Si}{i},
      and open circles are \ion{Ca}{i}.  Error bars for single elements removed for clarity. Filled circles with error bars
      represent \emph{weighted} means of Mg, Si, and Ca. Increasing in metallicity from left we have \object{EIS 7873},
      30077 and 6631.  } \label{figalpha}
\end{figure}
\begin{figure}
   \centering
   \includegraphics[width=6.5cm]{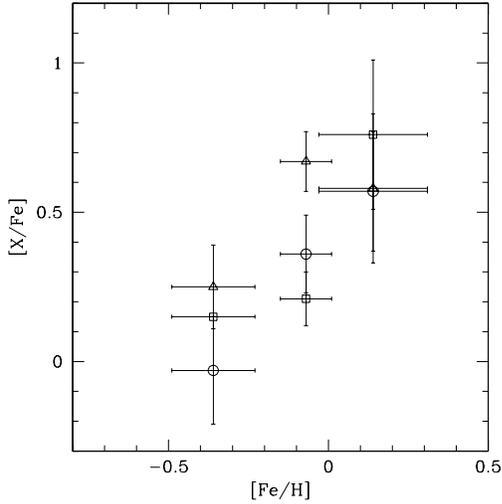}
      \caption{ Abundances for n-capture elements La (triangles),
 Ce (squares) and Nd (circles) plotted against metallicity.
              }
         \label{figenne}
   \end{figure}
\begin{figure}
   \centering
   \includegraphics[width=7.0cm]{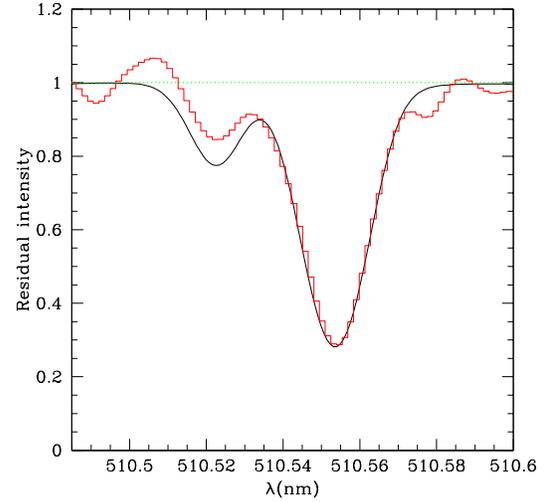}
      \caption{ Synthesis for the \ion{Cu}{i} 510.5 nm line in \object{EIS 6631} (thick line)
with superimposed the observed spectrum.
      Hyperfine splitting is taken into account, [Cu/Fe]=+0.25 in the synthesis.    }
         \label{figcu}
   \end{figure}

The atomic  data  and  lines used are  essentially   the ones used  in
\citet{bonifacio00}, the line list for iron  and $\alpha$ elements has
been expanded,  and new  $\log(gf)$ used  for \ion{Mg}{i}
\citep{gratton03}, \ion{Ca}{i} \citep{smith81},  \ion{La}{ii} \citep{lawler01} \ion{Nd}{ii}
\citep{hartog03} and \ion{Eu}{ii}  \citep{lawler01b}. The equivalent  widths
and atomic data used are available on request.  Derived abundances are
listed in Table \ref{tabdue}.

\section{Discussion and conclusions}

The  three stars appear to be  rather metal rich, ranging between $\rm
-0.5  \la  [Fe/H]\la +0.1$.  The   abundances of $\alpha$ elements are
shown  in Fig.  \ref{figalpha}  and  those of  n-capture elements   in
Fig. \ref{figenne}.  In order to  see whether the chemical  abundances
of these  three stars show any differences  from Galactic stars let us
compare with the   large  sample  compiled by \citet{venn04}.     This
catalogue contains  also a probability of  each star to  belong to the
Halo, Thick disc or  Thin disc. We selected only  stars for
which  the population membership  probability exceeds 85\%. The
mean values  and dispersion for  some significant abundance ratios
are given  in  Table \ref{tabvenn}.  From this   comparison we may see
that star \object{EIS 7873} appears to be undistinguishable from Galactic stars
of similar metallicity.  In this regime  the three components are very
similar, differing from each  other by not more than  1$\sigma$.
EIS    30077     and \object{EIS 6631},      instead, display    a  significant
underenhancement of    $\alpha   $ elements (except   for    Ti) and a
significant enhancement   over the  solar values of   La, Ce  and  Nd.
Unfortunately there is  very little data  available on  these elements
for  Galactic stars  in  this metallicity   regime.  The situation  is
better for Y, for this element \object{EIS 6631} shows [Y/Fe]$\approx 0$, while
\object{EIS 30077} shows  a  strong deficiency [Y/Fe]$\approx -0.6$;  note that
also Galactic stars display a large scatter in Y abundances.  \object{EIS 6631}
and \object{EIS  30077} 
display   another    remarkable abundance anomaly:    a
significant   overabundance  of  Cu   ([Cu/Fe]$\approx  +0.25$).  This
abundance is based on a single line, shown in Fig. \ref{figcu} for EIS
6631.     The  line appears clean    and  the fitting straightforward,
although all  the usual {\it caveats}  on  speculating on an abundance
derived from a single line apply.  Galactic stars in the
high metallicity regime show [Cu/Fe]$\sim 0$ \citep{bihain}. Moreover,
this resemblance strengthens the hypothesis  that these two stars have
a common origin.

\begin{table}
\begin{center}
\caption{Abundance ratios in different Galactic components.}
{\scriptsize
\begin{tabular}{lrrrrr}
\hline\\
          & (1)  & (2) & (3) & (4) & (5)\\
\hline\\
N       &  16  & 11 & 35 & 31 & 4 \\
$\rm [\alpha/Fe]        $& 0.14 & 0.06 & 0.06 & 0.02 & 0.08 \\
$\rm [\alpha/Fe]_\sigma $& 0.09 & 0.03 & 0.04 & 0.03 & 0.05\\
$\rm [Na/Fe]            $& 0.06 & 0.03 & 0.05 & 0.00 & --\\
$\rm [Na/Fe]_\sigma     $& 0.04 & 0.08 & 0.04 & 0.07 & --\\
$\rm [Y/Fe]             $& 0.00$^a$ & $-0.11^b$& 0.11$^c$& 0.04 & $-0.06$ \\
$\rm [Y/Fe]_\sigma      $& 0.24 & 0.14 & 0.58 & 0.10 & 0.03 \\

\hline\\
\multicolumn{6}{l}{(1) Thick disc ~ $\rm -0.5\le [Fe/H]< -0.2$}\\
\multicolumn{6}{l}{(2) Thick disc ~ $\rm-0.2 \le [Fe/H]$}\\
\multicolumn{6}{l}{(3) Thin disc ~$\rm -0.5\le [Fe/H]< -0.2$}\\
\multicolumn{6}{l}{(4) Thin disc ~$\rm-0.2 \le [Fe/H]$}\\
\multicolumn{6}{l}{(5) Halo ~ $\rm -0.5\le [Fe/H] $ }\\
\multicolumn{6}{l}{$^a$ 8 stars;$^b$ 5 stars; $^c$ 28 stars }\\
\end{tabular}
}
\label{tabvenn}
\end{center}
\end{table}

It is also interesting to compare these abundances with those of Local
Group dwarf galaxies.  Among these the only one which has a population
as metal--rich  as   our stars is    Sagittarius \citep[and references
therein]{bonifacio04}.   Sagittarius  is   characterized  by  a    low
[$\alpha$/Fe] and in this respect it  is similar to  
\object{EIS 30077} and \object{EIS 6631}.
This characteristic   is  shared by    other dwarf  spheroidal
galaxies in   the    LG \citep{shetrone04,venn04},   which  are   more
metal--poor than Sgr. This feature is  generally interpreted as due to
a low    star formation   rate  in  these  galaxies.   Therefore   the
[$\alpha$/Fe] ratios support  the notion that  
\object{EIS 30077}  and \object{EIS 6631}
have  not been  formed in   the Galaxy   but   rather in  a  LG  dwarf
spheroidal.  Another     ``signature''  of Sgr is  a     rather strong
overabundance in heavy neutron capture elements La, Ce, Nd and Eu; EIS
30077 and \object{EIS 6631} seem to behave in the  same way.  On the other hand
Sgr displays low [Na/Fe], [Ni/Fe], [Mn/Fe] and [Cu/Fe] ratios
\citep{bonifacio00,mcw03} while we find solar
ratios for our stars (except for Cu).

Field contamination may constitute  a significant issue.  By comparing
the derived  heliocentric velocity distribution of  our GIRAFFE target
stars to Galactic models we  found that contamination by disc stars is
present at any radial velocity \citep{zaggia04}, thus radial velocity,
by itself, does not ensure membership to the structure.
We need
both radial velocity and
metallicity to  isolate  possible  CMa
members. 
Our  estimate for the  mean heliocentric V$\rm_{rad}$ of  CMa in
the  background of \object{NGC  2477} is  of about  $132.0$ \kms\
with  a velocity  dispersion  of  $\simeq12.0$ \kms  for   a sample  of
$\simeq 20$  stars.
The resolution of our GIRAFFE spectra is $\sim$ 18 \kms,
which allows us to measure radial velocities with an internal
accuracy of the order of 1 \kms.
While  all three the    stars in Table \ref{tabuno}  have a  radial
velocity  within 3  $\sigma$  of this,  only  \object{EIS 6631}  falls
inside  the  1 $\sigma$  boundary.
\object{EIS  6631},  our ``best  candidate'',
appears also  to be  the most $\alpha-$poor  and the most  enhanced in
n-capture elements.

Using  the data  compiled   by \citet{venn04} we  find that
negative [$\alpha$/Fe] ratios are observed in only three disc stars at
such  high metallicities, the most  $\alpha-$poor being  HR 7126 with
[$\alpha$/Fe]=--0.08.
Thus the underenhancement of \object{EIS 6631} seems to be
rather unique   and constitutes    a   fairly  strong case   for    an
extragalactic origin.  Also the  significant Cu overabundance in this  star,
taken at face value, suggests an extra-galactic origin.

The conclusions that may  be drawn from our  observations are not very
compelling.
\object{EIS 7873}  appears to be  undistinguishable from Galactic
stars.    For \object{EIS  30077}  and
\object{EIS  6631} there   are  some clues for an
extra-galactic origin,  and these are  stronger for \object{EIS 6631},
the most metal-rich star of the  sample. This is surprising, since CMa
is  a highly  ``degraded'' structure,  at  variance with the \object{Sgr dSph},
most of the (hypothetical)  galaxy has already  dissolved, and its gas
content, whatever it may have been, has likely  mixed with that of the
MW.  Colours
and metallicity of \object{EIS 6631} imply a young age (about 2
Gy according to isochrones). It should then have formed
from gas in  which possible chemical signatures  may have been already
diluted.  If \object{EIS 6631} actually belongs to an external galaxy, its high
metallicity requires that the  mass of this galaxy  should be as large
as that of Sgr($M\ge10^9 M_\odot$  \citealt{ibata97}), or larger. This
is consistent  with the high end of  the mass estimate  of the Ring by
\citet{ibata}  ($2\times 10^8M_\odot\le  M\le 10^9$).   Larger samples
are required in order to  shed more light  on the origin and nature of
the CMa overdensity.

\begin{acknowledgements}
We are grateful to J.E. Lawler for providing
the HFS data for \ion{Eu}{ii} in machine readable form.
We  acknowledge financial support of MIUR (COFIN 2002028935\_003).
\end{acknowledgements}

\end{document}